\documentclass[letterpaper,english,reprint,aps,pra,superscriptaddress,notitlepage,onecolumn,nofootinbib]{revtex4-1}
\usepackage[T1]{fontenc}
\usepackage[latin9]{inputenc}
\usepackage{xurl}
\PassOptionsToPackage{hyphens}{url}\usepackage{hyperref}
\usepackage{color}
\usepackage{verbatim}
\usepackage{amsmath,braket}
\usepackage{amsthm,bm}
\usepackage[normalem]{ulem}
\usepackage{amssymb}
\usepackage{graphicx}
\usepackage{subfigure}
\newtheorem{lemma}{Lemma}
\newtheorem{Pro}{Proposition}
\newtheorem{theorem}{Theorem}
\newtheorem{Cor}{Corollary}

\makeatletter
\def\<{\langle}
\def\>{\rangle}

\pdfpageheight\paperheight
\pdfpagewidth\paperwidth

\usepackage{graphicx}
 \usepackage{hyperref}
\usepackage{pgfplots}
\usepgfplotslibrary{fillbetween}

\usepackage{listings}
\lstdefinestyle{empty}{
  backgroundcolor=\color{white},
  language=,
  numbers = none,
  frame=none,
}
\usepackage{tikz}
\usetikzlibrary{patterns}
\usepackage{pgfplots}
\usetikzlibrary{positioning}
\usetikzlibrary{decorations.markings}%a
\usetikzlibrary{arrows,decorations.text}
\usetikzlibrary{backgrounds,fit,decorations.pathreplacing}

\usepackage{babel}

\newtheorem{remark}{Remark}

\makeatother

\usepackage{babel}
\begin{document}
	\title{Generalizing Choi map in $M_3$ beyond circulant scenario}
	%On positive  maps in the matrix algebra $M_3$
 \author{Anindita Bera}
	\affiliation{Institute of Physics, Faculty of Physics, Astronomy and Informatics,
		Nicolaus Copernicus University, Grudziadzka 5/7, 87-100 Toru\'{n},
		Poland}
	\author{Giovanni Scala}
	\affiliation{International Centre for Theory of Quantum Technologies (ICTQT), University
		of Gdansk, Wita Stwosza 63, 80-308 Gda\'{n}sk, Poland}
	\affiliation{Faculty of Physics, University of Warsaw, Pasteura 5, 02-093 Warsaw, Poland}
	\author{Gniewomir Sarbicki}
	\affiliation{Institute of Physics, Faculty of Physics, Astronomy and Informatics,
		Nicolaus Copernicus University, Grudziadzka 5/7, 87-100 Toru\'{n},
		Poland}
	
	\author{Dariusz Chru\'{s}ci\'{n}ski}
	\affiliation{Institute of Physics, Faculty of Physics, Astronomy and Informatics,
		Nicolaus Copernicus University, Grudziadzka 5/7, 87-100 Toru\'{n},
		Poland}
	\date{\today}
	\begin{abstract}
		\noindent We present a generalization of the family of linear positive maps in $M_3$  proposed thirty years ago by Cho {\em et al.} (Linear Algebra Appl. {\bf 171}, 213 (1992)) as a generalization of the seminal Choi non-decomposable map. The necessary and sufficient conditions for decomposability are provided.

	\end{abstract}
	\maketitle
	
	\section{Introduction}

Positive linear maps in matrix algebras play important role both in mathematics and physics \cite{Stormer-63,Stormer,Paulsen,Tomiyama1,Tomiyama2,Tomiyama3,Bhatia,HHHH,Guhne,TOPICAL}. Let $M_n$ denote a $C^*$-algebra of $n\times n$ complex matrices.  A linear map $\Phi : M_n\to M_m$ is positive if $\Phi(X) \geq 0$ for all $X \geq 0$. The structure of a convex cone ${\cal P}_{n,m}$ of positive maps from $M_n$ to $M_m$ is still not fully understood \cite{MM,OSID,CMP,KOREA,Ha-Kye,TOPICAL}. Due to Woronowicz \cite{Woronowicz} the cones  ${\cal P}_{2,2}$,  ${\cal P}_{2,3}$, and  ${\cal P}_{3,2}$ consist of decomposable maps only,  that is, maps which can be  decomposed as follows  

\begin{equation}
    \Phi = \Phi_1 + \Phi_2\circ {\rm T} ,
\end{equation}
where $\Phi_1,\Phi_2$ are completely positive, and ${\rm T}$ denotes a matrix transposition. Recall that a positive map $\Phi \in {\cal P}_{n,m}$ is completely positive if the extended map 
\begin{equation}
\Phi^{(k)} := {\rm id}_k \otimes \Phi : M_k(M_n) \to M_k(M_m) ,    
\end{equation}
is positive for all $k=1,2,\ldots$ (${\rm id}_k$ denotes an identity map in $M_k$).  Actually, due to \cite{Choi-75}, the map $\Phi$ is complete positive if $\Phi^{(k)}$ is positive for $k = {\rm min}\{n,m\}$. A first example of a non-decomposable map in $\mathcal{P}_{3,3}$ was provided by Choi \cite{Choi1,Choi2,Choi3}
\begin{equation}
\label{Choi1}
  \Phi(X) = \begin{bmatrix}
   x_{11} + x_{22} &-x_{12} & -x_{13} \\ -x_{21} & x_{22}+x_{33} & - x_{23} \\ -x_{31} & -x_{32} & x_{33}+x_{11}
\end{bmatrix},
\end{equation}
with $X =\{x_{ij}\} \in M_3$. Choi map (\ref{Choi1}) was then generalized to a 3-parameter family of maps \cite{Korea1}
\begin{equation}   \label{Korea}
\Phi_C (X) = D(X) - X  ,    
\end{equation}
where $D(X)$ is a diagonal matrix with $D_{ii}= \sum_{j=1}^3 C_{ij} x_{jj}$, and $C=\{C_{ij}\}$ is a $3\times 3$ circulant matrix in the following form
\begin{equation}
\label{abc}
C = \begin{bmatrix}
 a&b&c    \\ c&a&b \\ b&c&a
\end{bmatrix} , \ \ \ a,b,c \geq 0 .
\end{equation}
For $(a,b,c) = (2,1,0)$ the map (\ref{Korea}) reduces to (\ref{Choi1}). The authors of \cite{Korea1} proved the following:

\begin{theorem} \label{TH-Korea} The map $\Phi_C$ defined in (\ref{Korea})  is positive if and only if
\begin{itemize}
    \item $a \geq 1$, 
    \item  $a + b + c \geq 3$,
    \item if $a \leq 2$, then $bc \geq (2-a)^2$.
\end{itemize}
For $ a < 3$, the positive map $\Phi_C$ is decomposable if and only if 

\begin{equation}\label{DEC}
    bc \geq  \left(\frac{3-a}{2}\right)^2 .
\end{equation}
If $a \geq 3$, then $\Phi_C$ is completely positive (and hence decomposable).  
%being positive it is non-decomposable if and only if $bc < (2-a)^2$. 
%The map (\ref{Korea}) is   completely positive if and if $a \geq 3$.
\end{theorem}
In this paper, we generalize the above result for a class of maps defined in (\ref{Korea}), where now $D(X)$ is a diagonal matrix with $D_{ii}= \sum_{j=1}^3 w_{ij} x_{jj}$, and $W = \{w_{ij}\} \in M_3$ such that $w_{ij} \geq 0$
and

\begin{equation}  \label{ww}
    \sum_{i=1}^3 w_{ij} = \sum_{j=1}^3 w_{ij} = w ,
\end{equation}
i.e. $W/w= \{w_{ij}/w\}$ is a doubly stochastic matrix. In what follows a map defined in terms of the matrix $W$ by $\Phi_W$. It is clear that when $W=[w_{ij}]$ is circulant, then our class reduces to the from \eqref{Korea}. 

This paper is organized as follows. In Sec. \ref{main-result}, we derive the conditions for positivity of the map $\Phi_W$. In Sec. \ref{new-para}, we present a new parameterization for $W$ from Birkhoff theorem and analyze the obtained family of positive maps. The necessary and sufficient condition for decomposability  is discussed in Sec. \ref{decnew}. Finally, in Sec. \ref{conclude}, we provide concluding remarks.

	\section{positivity}
	\label{main-result}
Let us consider a linear map $\Phi_W$ 	parameterized by a $3 \times 3$ matrix $W$ with non-negative entries $\{w_{ij}\}$. Note that $\Phi_W$ is completely positive if and only if

\begin{equation}\label{CP}
    \left(\begin{array}{ccc}
		w_{11}-1 & -1 & -1\\
		-1 & w_{22}-1 & -1 \\
		-1 & -1 & w_{33}-1
	\end{array}\right) \geq 0.
\end{equation}
The above condition is equivalent to
\begin{equation}
\label{CP1}
    \frac{1}{w_{11}} + \frac{1}{w_{22}} + \frac{1}{w_{33}} \leq 1 .
\end{equation}
It means that the harmonic mean
\begin{equation}
   H({w_{11}},{w_{22}},{w_{33}}) \geq 3 .
\end{equation}
In particular, if all $w_{ii}=a$, then (\ref{CP}) is equivalent to $H(a,a,a) = a \geq 3$. Now we find the conditions upon $W$ which guarantee that $\Phi_W$ is positive. It is enough to check the positivity for rank-one positive matrices $X=\psi\psi^\dagger$:% with $\psi_i$ components:
	\begin{equation} \label{pos1}
	    \forall \psi \in \mathbb{C}^3, \quad \Phi_W(\psi\psi^\dagger) = \mathrm{diag} \left( \sum_{i=1}^3 w_{1i} |\psi|^2_i, \sum_{i=1}^3 w_{2i} |\psi|^2_i, \sum_{i=1}^3 w_{3i} |\psi|^2_i \right) -\psi\psi^\dagger \ge 0.
	\end{equation}
	Let us denote $x_i = |\psi_i|^2$, $z_i = \sum_{j=1}^3 w_{ij} x_j$.
	The principal minor $M_I$ of $\Phi_W(\psi\psi^\dagger)$, constructed from the rows and columns of indices from the set $I \subset \{1,2,3\}$ is equal to
	\begin{equation}
        M_I (\vec x) = \prod_{i \in I} z_i - \sum_{i \in I} x_i \prod_{j \in I \setminus \{ i \}} z_j,
        \qquad\mbox{with  } \prod_{j \in \emptyset} z_j=1.
	\end{equation}
	Observe that, the positivity depends only on modules of $x_i$, and not on their phases. Moreover, the norm of $\psi$, equals to $x_1 + x_2 + x_3$, is irrelevant for positivity. Hence one can rewrite the condition (\ref{pos1}) as:
	\begin{equation} \label{minors}
	    \forall \vec x \in \Delta^2,~ \ \forall I \subset \{1,2,3\},~ \ M_I(\vec{x}) \ge 0, 
	\end{equation}
	where $\Delta^2$ denotes the 2-simplex with vertices $x_i$.
 Observe that, at most one eigenvalue of $\Phi_W(\psi\psi^\dagger)$ can be negative since it is obtained as a semi-positive matrix minus rank-one semi-positive matrix, hence in the generic case it is enough to check its determinant, i.e. $M_{\{1,2,3\}}$. Specifically, when $x_i = 0$ and the corresponding $z_i = 0$ might not be straightforward to determine the positivity. Then, the determinant can be zero, but the matrix $\Phi_W(\psi\psi^\dagger)$ can have a negative diagonal block. Therefore when some components of $x_i$ are zero and the $\Phi_W(\psi\psi^\dagger)$ obtains the block-diagonal structure, the solution is to analyze the positivity of the maximal non-trivial block.
	
	Proceeding in this way, we will check the above condition on disjoints subsets of $\Delta^2$ emerging from its cellular decomposition, i.e. on the interior, on the edges and in the vertices.
	
In the vertices of $\Delta^2$ (where any two components of $x_i$'s are zero), the positivity of $M_{\{1,2,3\}} (\vec x)$ reduces to positivity of size-1 minors and one obtains:
	\begin{equation} \label{vert1}
	    w_{ii} \ge 1,\qquad i=1,2,3.
	\end{equation}
	
In an edge of $\Delta^2$  (where any one component of $x_i$ is zero), let's say for $x_3 = 0$, the positivity of $M_{\{1,2,3\}} (\vec x)$ reduces to positivity of $M_{\{1,2\}} = z_1 z_2 - x_1 z_2 - x_2 z_1$:
	\begin{equation} \label{edge_x}
        (w_{11}-1)w_{21} x_1^2 + (w_{22}-1)w_{12}x_2^2 + [(w_{11}-1)(w_{22}-1)-1+w_{12}w_{21}] x_1 x_2 \ge 0.
	\end{equation}
	Using (\ref{vert1}), the above quadratic form is non-negative for positive $x_1,x_2$ iff $((w_{11}-1)(w_{22}-1)-1+w_{12}w_{21})> - 2 \sqrt{(w_{11}-1)(w_{22}-1)w_{21}w_{12}}$, which can be simplified as
    \begin{equation} \label{edge}
     \sqrt{(w_{11}-1)(w_{22}-1)} + \sqrt{w_{12}w_{21}} \ge 1.
	\end{equation}
	Similar conditions arise from the remaining edges. Summarizing the edge conditions read
	
 \begin{equation} \label{edge-1}
     \sqrt{(w_{ii}-1)(w_{jj}-1)} + \sqrt{w_{ij}w_{ji}} \ge 1 , 
	\end{equation}
for $i \neq j$. Note that, for $i=j$, condition (\ref{edge-1}) is equivalent to the vertex condition (\ref{vert1}). 
	
In the interior of $\Delta^2$, denoting as $\mathrm{Int} \Delta^2$  (where all $x_i >0$), the positivity of $M_{\{1,2,3\}}$ gives the third order polynomial inequality of $x_1,x_2,x_3$:
	\begin{equation} \label{inter1}
	    \forall \vec x \in \mathrm{Int} \Delta^2, \ z_1 z_2 z_3 - x_1 z_2 z_3 - x_2 z_3 z_1 - x_3 z_1 z_2 \ge 0.
	\end{equation}
	We want to find the values of local maxima of the LHS in (\ref{inter1}). Substituting $x_3 = 1- x_2 - x_1$ and demanding partial derivatives with respect to $x_1,x_2$ to be zero, one obtains a system of two second order equations, which have in general four solutions for extrema, arising from roots of a four-order polynomial. Hence, in general, the problem of finding the local maxima for `$z_1 z_2 z_3 - x_1 z_2 z_3 - x_2 z_3 z_1 - x_3 z_1 z_2$' is quite involve. Note that, condition (\ref{inter1}) may be equivalently rewritten as follows
	
		\begin{equation}\label{eq:fx}
	    f(\vec x) = \sum_{i=1}^3 \frac{x_i}{z_i} \le 1,
	\end{equation}
	where we assumed $z_1, z_2, z_3 > 0$. To find the maximum of the function $ f(\vec x)$ within the simplex $\mathrm{Int} \Delta^2$, one has to analyze the property of the corresponding Hessian matrix. One finds

\begin{equation}
    \frac{\partial^2 f(\vec x)}{\partial x_i \partial x_j} = - \hat{W}_{ij} ,
\end{equation}
where the matrix $\hat{W}$ is defined as follows

\begin{equation}\label{hat-W}
	    \hat W=w(W + W^T) - 2 W^TW .
	\end{equation}
It turns out that the problem simplifies when the matrix $W$ satisfies (\ref{ww}), that is, it is (up to a factor $w$) a doubly stochastic matrix. Then the function $f(\vec x)$ has  a unique extremum attained at $x_1=x_2=x_3$. This is due to the following  theorem \cite{Nowosad,Yamagami}:
	
\begin{theorem} \label{novisaarise}

	Let $W=[w_{ij}]$ be an invertible $3 \times 3$ matrix with non-negative real entries satisfying (\ref{ww}).	
	If the matrix $\hat{W}$ defined in (\ref{hat-W}) is positive semidefinite and its kernel is spanned by $\bm{1}=(1,1,1)$, then $\bm{1}$ is a unique (up to scalar factor) point that gives a local maximum inside the simplex  $\mathrm{Int}\Delta^2$ of the function $f$ defined in (\ref{eq:fx}).
	\end{theorem}
Actually, authors of \cite{Nowosad,Yamagami} formulated the above Theorem for arbitrary dimension $n$.

\iffalse
\textcolor{green}{
\begin{remark}
In such case, 	the local maximum is also a global one.
\end{remark}
Here $\bm{1}$ is a unique local maximum of $f(x_i)$ (=LHS of Eq.~(\ref{eq:fx}). If $W$ has the block-diagonal structure, then the problem simplifies from 3D to 2D which is easy to solve. However, in our paper, $W$ has no  block-diagonal structure and in the limiting case, the function  $f(x_i)$ is still continuous, hence there is no singular boundary and we can proceed with our formulation of edge and vertex conditions. 
}
\fi

\begin{remark} Note that when the off-diagonal elements $w_{12}$ or $w_{13}$ are different from zero then condition (\ref{eq:fx}) 

\begin{equation}
\label{c1}
    \frac{x_1}{w_{11} x_1 + w_{12} x_2 + w_{13}x_3} +  \frac{x_2}{w_{21} x_1 + w_{22} x_2 + w_{23}x_3} +  \frac{x_3}{w_{31} x_1 + w_{32} x_2 + w_{33}x_3} \leq 1 ,  
\end{equation}
reduces for $x_1 \to 0 $ to

\begin{equation}
      \frac{x_2}{ w_{22} x_2 + w_{23}x_3} +  \frac{x_3}{ w_{32} x_2 + w_{33}x_3} \leq 1 ,  
\end{equation}
which recovers the edge condition (\ref{edge-1}) for $i=2$ and $j=3$. However, when $w_{12}=w_{13}=0$, one obtains from \eqref{c1}, in the limit $x_1 \to 0 $ the following condition

\begin{equation}
     \frac{1}{w_{11}} + \frac{x_2}{ w_{22} x_2 + w_{23}x_3} +  \frac{x_3}{ w_{32} x_2 + w_{33}x_3} \leq 1 . 
\end{equation}
Such a scenario corresponds to the so-called {\em singular boundary} \cite{Yamagami}. This, however, corresponds to a block-diagonal matrix

\begin{equation}
W = \begin{bmatrix}
 w_{11}&0&0    \\ 0&w_{22}&w_{23} \\ 0&w_{32}&w_{33}
\end{bmatrix} .
\end{equation}
In what follows we assume that the $W$ matrix is not block diagonal. 
\end{remark}	

Supremum of the function $f$ on $\mathrm{Int}\Delta^2$ defines, therefore,  the maximum of the unique local maximum and the maximal value of the boundary and it cannot exceed one. The restriction of the boundary values is already taken into consideration in the Eqs.  (\ref{vert1}) and (\ref{edge-1}). 
Note that, condition (\ref{eq:fx}) reduces for $x_1=x_2=x_3$ to 
	\begin{equation} \label {inter2}
	    \frac 3{w_{11} + w_{12} + w_{13}} \le 1 ,
	\end{equation}
or, equivalently, $w \geq 3$. 

\begin{lemma} If $W$ satisfies (\ref{ww}), then $|w_{ij} - w_{ji}|$ does not depend upon $i \neq j$.
\end{lemma}
The proof is straightforward. Let us define

\begin{equation} \label{delta}
    \delta := |w_{ij} - w_{ji}|  ,  \ \ \ i\neq j .
\end{equation}
Note that, the above property holds for $3\times 3$ matrices only. In particular, if $W_{ij}=C_{ij}$ is circulant, then 

\begin{equation}
    \delta = |b-c| .
\end{equation}

\begin{lemma} The matrix $\hat W=w(W + W^T) - 2 W^TW \geq 0$ if and only if

\begin{equation}  \label{www}
\sum_{i=1}^3 (w_{ii}-w_{i+1,i+1})^2 \leq \frac 12 \left( w - \sqrt{ ({\rm Tr}\, W - 2 w)^2 + 3\delta^2} \right)^2 ,    
\end{equation}
where $\delta$ is defined in (\ref{delta}), and the summation on the LHS is mod 3.  
\end{lemma}

\begin{theorem} \label{th1} Let $W$ be a $3\times 3$ matrix with non-negative elements satisfying (\ref{ww}).  If moreover $W$ satisfies  (\ref{www}), then the linear map $\Phi_W$ is positive if and only if 

\begin{itemize}
    \item $w_{ii} \geq 1$ for $i=1,2,3$ ({\em vertex conditions})
    \item  $\sqrt{(w_{ii}-1)(w_{jj}-1)} - \sqrt{w_{ij}w_{ji}} \geq 1$, for $i\neq j$  ({\em edge conditions})
    \item $w\geq 3$ ({\em interior condition}).
\end{itemize}
	    
\end{theorem}
Note that, if $w_{ij}=C_{ij}$  is cirulant, then condition (\ref{www}) is trivially satisfied, and then  Theorem \ref{th1} reduces to Theorem \ref{TH-Korea}.  

\section{New parameterization from Birkhoff theorem}
\label{new-para}

Since $W/w$ is a doubly stochastic matrix, then due to the Birkhoff theorem, it can be represented as a convex combination of the permutation matrices, that is,

\begin{equation}
    W = 
    \begin{bmatrix}
 a & b & c    \\ c & a & b \\ b & c & a
\end{bmatrix} 
+ d 
\begin{bmatrix}
 0 & 1 & 0    \\ 1 & 0 & 0 \\ 0 & 0 & 1
\end{bmatrix} 
+ e 
\begin{bmatrix}
 0 & 0 & 1    \\ 0 & 1 & 0 \\ 1 & 0 & 0
\end{bmatrix}
+ f 
\begin{bmatrix}
 1 & 0 & 0    \\ 0 & 0 & 1 \\ 0 & 1 & 0 
\end{bmatrix}  
\nonumber 
 = \left(\begin{array}{ccc}
		a+f & b+d & c+e\\
		c+d & a+e & b+f\\
		b+e & c+f & a+d
	\end{array}\right) ,
\end{equation}
with non-negative $\{a,b,c,d,e,f\}$ satisfying $a+b+c+d+e+f=w$.
Clearly, under this parameterization, one can rewrite the condition of complete positivity in \eqref{CP1} in the below form:
\begin{equation} \label{CP2}
    \frac{1}{a+d}+\frac{1}{a+e}+\frac{1}{a+f} \leq 1.
\end{equation}

Note that the parameters $\{d,e,f\}$ perturb the circulant part $C_{ij}$. However, this representation is not unique. Actually, performing the following {\em gauge} transformation

\begin{equation}
    \{a,b,c\} \to \{a+\xi,b+\xi,c+\xi\} \ , \ \ \ \{d,e,f\} \to \{d-\xi,e-\xi,f-\xi\} ,
\end{equation}
one does not affect the matrix elements $w_{ij}$'s (which have to be non-negative). Clearly, the normalization condition $a+b+c+d+e+f=w$ is gauge invariant. In what follows we fix the gauge condition

\begin{equation} \label{gauge}
    d+e+f=0 .
\end{equation}
Notice that in such a gauge ${\rm Tr}\, W = 3a$, $w = a+b+c$, and $\delta = |b-c|$. 
Moreover, if $S$ is a permutation defined by $Se_i = e_{i+1}$ (mod 3), then the circulant part $C$ of $W$ satisfies

\begin{equation}   \label{CW}
    C = \frac 13(W + S W S^{\rm T} + S^{\rm T} W S ) , 
\end{equation}
that is, {\em averaged} matrix $W$ is circulant.

\begin{Pro} \label{Pro-1} If $\Phi_W$ is a positive map, then $\Phi_C$, with $C$ being a circulant part of $W$ defined in  (\ref{CW}) is also  positive.
\end{Pro}
Proof: note that due to (\ref{CW}) one has (cf. also \cite{Vietnam})

\begin{equation}
    \Phi_C(X) = \frac 13 \left( \Phi_W(X) +   S \Phi_W(S^{\rm T} X S) S^{\rm T} + S^{\rm T} \Phi_W(SXS^{\rm T}) S \right) , 
\end{equation}
and hence $\Phi_C$ is a convex combination of three positive maps. \hfill $\Box$

The above proposition provides a simple necessary condition for positivity of the map $\Phi_W$. 

Note that, using (\ref{gauge}) the condition (\ref{www}) can be rewritten as follows
\begin{equation} \label{hess2}
	        (d-e)^2 + (e-f)^2 + (f-d)^2 \le 
	        %p + q + \sqrt{(p+q)^2 - p^2},
	        \frac 12 \left( a+b+c - \sqrt{(a-2b-2c)^{2}+3(b-c)^{2}} \right)^2 .
	    \end{equation}
%Note, that
The vertex condition is $w_{ii}\geq 1$ and $w_{ij} \geq 0$ for $i \neq j$ if

\begin{equation} \label{vert}
	        d,e,f \ge -\min\{a-1,b,c\} \stackrel{df}= -\mu .
	    \end{equation}
	 Edge conditions have the following form
\begin{align}
	        \sqrt{(a+f-1)(a+e-1)} + \sqrt{(b+d)(c+d)} \ge 1, 
	        \label{edge1} \\
	       \sqrt{(a+d-1)(a+f-1)} + \sqrt{(b+e)(c+e)} \ge 1,
	       \label{edge2} \\
	        \sqrt{(a+e-1)(a+d-1)} + \sqrt{(b+f)(c+f)} \ge 1, 
	        \label{edge3}
	    \end{align}
and the interior one
	    \begin{equation}
	        a + b + c \ge 3. \label{interior}
        \end{equation}

    Under such  characterization,  we analyze the obtained family of positive maps. We describe it in its space of parameters $\{a,b,c\}$ on the plane $d+e+f = 0$. By a closer inspection, condition \eqref{hess2} describes a circle of the radius  $\frac 1{\sqrt{6}}\left( a+b+c - \sqrt{(a-2b-2c)^{2}+3(b-c)^{2}} \right)$ on the plane $d+e+f=0$, centered in the origin. 
    
    Note also that, if $a \ge 1, b,c \ge 0$ then the condition (\ref{vert}) cuts off an equilateral triangle on the plane $d+e+f = 0$. Its vertices are $(2\mu,-\mu,-\mu)$, $(-\mu,2\mu,-\mu)$, $(-\mu,-\mu,2\mu)$ and we denote them $D$, $E$, $F$ respectively.
    
      The inequality (\ref{edge1}) describes a shape %a curve
    symmetric w.r.t. the height of the triangle passing through its vertex $D$. Similarly, the inequalities (\ref{edge2}) and (\ref{edge3}) describe identical shapes, but symmetric w.r.t. the heights passing through vertices $E$ and $F$ respectively.
    The intersection of these three shapes is then a curved equilateral triangle, whose sides are arcs defined by saturations of the inequalities \eqref{edge1}-\eqref{edge3}.  Each arc is an arc of a 4th-order bean curve \cite{Cundy}. 
    
    %and connected in angles $\pm \pi/3$, $\pi$. 
   In the special case of $b=c$, all the bean curves reduce to a circle centered in the point $d=e=f=0$.
  
    The intersection of the arcs is non-empty iff the point $d=e=f=0$ satisfies \eqref{edge1} - \eqref{edge3}, which yields:
    \begin{equation}
        2-a \le \sqrt{bc}.
    \end{equation}
    In this way, we recover theorem \ref{TH-Korea} for  the map $\Phi_C$ ($d=e=f=0$).
    
\begin{figure}[t]
        \centering
        \includegraphics[width=.32\textwidth]{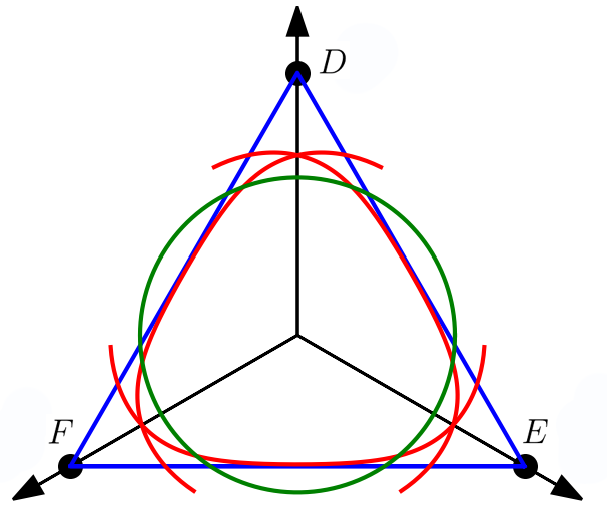} 
        \caption{A typical shape of the set of admissible points in the $d+e+f=0$ plane is a intersection of three configurations: a circle (green), a curved-arcs triangle (red) and a line-segments triangle (blue). Points outside the intersection of the triangles refer to non-positive maps. The positivity of maps referring to points in the intersection of the triangles, but outside the circle are not known. Here the values of the  parameters are $a=1.7,~b=0.9,~c=0.5$. The interactive graph varying the parameters $\{a,b,c\}$ is available on this link 
        \href{https://github.com/gniewko-s/positive_3x3_maps_bistochastic_on_diagonal/blob/main/equilateral.py}{https://github.com/gniewko-s/positive\_{$3\times 3$}\_maps\_bistochastic\_on\_diagonal/blob/main/equilateral.py}.
      }
        \label{fig:triangles}
    \end{figure}

    In the space of parameters $\{d,e,f\}$ of Fig. \ref{fig:triangles}, the typical shape of the set of admissible points in the plane $d+e+f=0$ is an intersection of the circle (hessian condition (\ref{hess2})), the curved-arcs triangle (edge conditions \eqref{edge1}--\eqref{edge3}) and the line-segments triangle (vertex conditions \eqref{vert}), mentioned in green, red and blue colors, respectively. See Appendix~\ref{appA} for the  numerical details of Fig. \ref{fig:triangles}.

\section{Decomposability}
\label{decnew}

Let $E_{ij}$ denotes the matrix units in $M_3$. Given a linear map $\Phi : M_3 \to M_3$, due to the Choi-Jamio\l{}kowski isomorphism \cite{Jam}, one can write 
\begin{equation}
    \hat{\Phi} := \sum_{i,j=1}^3 E_{ij} \otimes \Phi(E_{ij}) \in M_3(M_3) .
\end{equation}
A linear positive map is decomposable if and only if
\begin{equation}
    \hat{\Phi} = A + ({\rm id}_3 \otimes {\rm T})B ,
\end{equation}
with $A,B \geq 0$. Now we provide (non-)decomposability conditions for the map $\Phi_W$. Note that proposition \ref{Pro-1} implies the following

\begin{Cor} If the map $\Phi_W$ is decomposable, then $\Phi_C$ is  also decomposable. Hence, whenever $\Phi_C$ is non-decomposable, then any $W$ compatible with $C$ (in the sense of (\ref{CW})) gives rise to a non-decomposable map $\Phi_W$. 
\end{Cor}

Now, we formulate a necessary and sufficient conditions for decomposability.

\begin{Pro} \label{Pro-2} Suppose that $\Phi_W$ is positive but not completely positive. If the following condition holds

\begin{equation}   \label{DEC-W}
    w_{ij} w_{ji} \geq \left( \frac{\sqrt{(w_{ii}-1)(w_{jj}-1)} - 2}{2} \right)^2 , \ \ \ i\neq j ,
  %  \langle \bm 1|\left(\begin{array}{ccc}
	%	w_{11} & \sqrt{w_{12} w_{21}} & \sqrt{w_{13} w_{31}}\\
	%	\sqrt{w_{21} w_{12}} & w_{22} & \sqrt{w_{23} w_{32}}\\
%		 \sqrt{w_{31} w_{13}} &  \sqrt{w_{32} w_{23}} & w_{33}
%	\end{array}\right)^{-1} |\bm 1 \rangle \leq 1,
\end{equation}
then the map $\Phi_W$ is decomposable. 
\end{Pro}
{\em Proof:} We can represent $\hat{\Phi}_W$ in the below form  (to make the structure of the matrices more transparent we replace $0$ by dots)

\begin{equation}   \label{}
    \hat{\Phi}_W = \left(
\begin{array}{ccc|ccc|ccc}
 w_{11}-1 & . & . & . & -1 & . & . & . & -1 \\
 . & w_{21} & . & . & . & . & . & . & . \\
 . & . & w_{31} & . & . & . & . & . & . \\  \hline
 . & . & . & w_{12} & . & . & . & . & . \\
 -1 & . & . & . & w_{22}-1 & . & . & . & - 1 \\
 . & . & . & . & . & w_{32} & . & . & . \\  \hline
 . & . & . & . & . & . & w_{13} & . & . \\
 . & . & . & . & . & . & . & w_{23} & . \\
 -1 & . & . & . & - 1 & . & . & . & w_{33} -1
\end{array}
\right) = A + ({\rm id}_3 \otimes {\rm T})B ,
\end{equation}
where the matrices $A$ and $B$ are defined by

\begin{equation}   \label{}
    A = \left(
\begin{array}{ccc|ccc|ccc}
 w_{11}-1 & . & . & . &  a_{12} & . & . & . & a_{13} \\
 . & . & . & . & . & . & . & . & . \\
 . & . & . & . & . & . & . & . & . \\  \hline
 . & . & . & . & . & . & . & . & . \\
  a_{21} & . & . & . & w_{22}-1 & . & . & . &  a_{23} \\
 . & . & . & . & . & . & . & . & . \\  \hline
 . & . & . & . & . & . & . & . & . \\
 . & . & . & . & . & . & . & . & . \\
 a_{31} & . & . & . &  a_{32} & . & . & . & w_{33} -1
\end{array}
\right) \ , \ \ \ \ 
B = \left(
\begin{array}{ccc|ccc|ccc}
 . & . & . & . & . & . & . & . & .\\
 . & w_{21} & . & b_{12} & . & . & . & . & . \\
 . & . & w_{31} & . & . & . & b_{13} & . & . \\  \hline
 . & b_{21} & . & w_{12} & . & . & . & . & . \\
 . & . & . & . & . & . & . & . & . \\
 . & . & . & . & . & w_{32} & . & b_{23} & . \\  \hline
 . & . & b_{31} & . & . & . & w_{13} & . & . \\
 . & . & . & . & . & b_{32} & . & w_{23} & . \\
 . & . & . & . & . & . & . & . & .
\end{array}
\right) ,
\end{equation}
with

\begin{equation}
    a_{ij} = - \frac 12 \sqrt{(w_{ii}-1)(w_{jj}-1)} , \ \ \ b_{ij} =  \frac 12 \sqrt{(w_{ii}-1)(w_{jj}-1)} -1 .
% a_{ij}=\sqrt{w_{ij} w_{ji}}-1, \ \ \ b_{ij}=-\sqrt{w_{ij} w_{ji}}.
 \label{aijbij}
\end{equation}
Observe that the conditions (\ref{DEC-W}) imply positivity of both $A$ and $B$. Indeed, $B \geq 0$ if and only if

\begin{equation}
w_{ij} w_{ji} \geq b_{ij} b_{ji} = b_{ij}^2 ,    
\end{equation}
which is equivalent to (\ref{DEC-W}). Positivity of $A$ follows immediately from the positivity of the following $3\times 3$ submatrix

\begin{equation}
    \left(\begin{array}{ccc}
		w_{11}-1 & a_{12} & a_{13}\\
		a_{21} & w_{22}-1 & a_{23}\\
		a_{31} & a_{32} & w_{33}-1
	\end{array}\right) ,
\end{equation}
which ends the proof. \hfill $\Box$

\begin{Pro} \label{Pro-3} A necessary condition for decomposability of $\Phi_W$ reads 
\begin{equation}   \label{NDEC-W}
\sum_{i=1}^3 w_{ii}+2 \sum_{i<j} \sqrt{w_{ij} w_{ji}} \geq 9 .
\end{equation}
\end{Pro}
Proof: recall that $\Phi_W$ is decomposable if and only if

\begin{equation}
    {\rm Tr}\, (\hat{\Phi}_W X) \geq 0 ,
\end{equation}
for any $X \in M_3(M_3)$ such that $X \geq 0$ together with  $({\rm id}_3 \otimes {\rm T})X \geq 0$. We  generalize  St{\o}rmer approach \cite{Stormer-DEC} and consider the following matrix $X \in M_3(M_3)$:

\begin{equation}
X =
\left(
\begin{array}{ccc|ccc|ccc}
 1 & . & . & . & 1 & . & . & . & 1 \\
 . & \epsilon _1 & . & . & . & . & . & . & . \\
 . & . & \frac{1}{\epsilon _2} & . & . & . & . & . & . \\ \hline
 . & . & . & \frac{1}{\epsilon _1} & . & . & . & . & . \\
 1 & . & . & . & 1 & . & . & . & 1 \\
 . & . & . & . & . & \epsilon _3 & . & . & . \\ \hline
 . & . & . & . & . & . & \epsilon _2 & . & . \\
 . & . & . & . &  & . & . & \frac{1}{\epsilon _3} & . \\
 1 & . &  & . & 1 & . & . & . & 1 \\
\end{array}
\right),
\end{equation}
with $\epsilon_i > 0$. By construction one has $X \geq 0$ and  $({\rm id}_3 \otimes {\rm T})X \geq 0$.
One finds

$$\mbox{Tr}(\hat{\Phi}_W X)= \sum_{i=1}^3 w_{ii} +(\epsilon _1 w_{21}+\frac{1}{\epsilon _1} w_{12}) +(\epsilon _2 w_{13}+\frac{1}{\epsilon _2} w_{31})+(\epsilon_3 w_{32}+\frac{1}{\epsilon_3} w_{23})- 9 , $$
and minimization w.r.t. $\{\epsilon_1,\epsilon_2,\epsilon_3\}$ implies (\ref{NDEC-W}). \hfill $\Box$ 

\begin{remark} Let us observe that due to $2\sqrt{w_{ij} w_{ji}} \leq w_{ij} + w_{ji} $  condition (\ref{NDEC-W}) implies

$$   \sum_{i} w_{ii} + \sum_{i<j}(w_{ij} + w_{ji}) = 3 w \geq 9 , $$
and hence $w \geq 3$. It is, therefore, clear that (\ref{NDEC-W}) is more restrictive than the interior condition $w \geq 3$. Note that, when $W=C$ (circulant case), then (\ref{DEC-W}) and (\ref{NDEC-W}) coincide and reduces to (\ref{DEC}). 
\end{remark}

\section{Conclusions}
\label{conclude}

In this paper, we generalized the family of positive maps proposed in \cite{Korea1}. The generalization consists in relaxing the circulant structure of the matrix $\{C_{ij}\}$, i.e. replacing $C_{ij}$ by $w_{ij}$ satisfying (\ref{ww}). Positivity of such maps is characterized in Theorem \ref{th1}. The corresponding sufficient and necessary conditions for decomposability are summarized in Propositions  \ref{Pro-2} and \ref{Pro-3}, respectively.  

It would be interesting to apply this new family of maps for the analysis of the detection of quantum entanglement, where e.g. the so-called optimality property plays a key role \cite{Lew,TOPICAL,KOREA,Ha-Kye,PRA-2022,ani1}.

\begin{acknowledgments}
		GSc is supported by QuantERA/2/2020, an ERA-Net co-fund in Quantum Technologies, under the eDICT project. GSc thanks the Institute of Physics of the Nicolaus Copernicus University of Toru\'n for the hospitality. AB, GSa and DC were supported by the Polish National Science Centre project No. 2018/30/A/ST2/00837.
	\end{acknowledgments}

\appendix

\section{Numerical calculations}
\label{appA}

The code \hyperlink{https://github.com/gniewko-s/positive_3x3_maps_bistochastic_on_diagonal/blob/main/equilateral.py}{\url{https://github.com/gniewko-s/positive\_3x3\_maps\_bistochastic\_on\_diagonal/blob/main/equilateral.py}} is written in the Python language and uses the \lstinline{numpy} and \lstinline{matplotlib} libraries. It draws on the plane $d+e+f=0$ the following curves being saturations of the following conditions:
\begin{itemize}
    \item vertex conditions (\ref{vert}) (the blue triangle)
    \item edge conditions (\ref{edge1} - \ref{edge3}) (the red curved-arcs triangle)
    \item the hessian positivity condition \ref{hess2} (the green circle)
    \item the complete positivity condition \ref{CP2} (the black shape)
\end{itemize}
The angles between projections of axes $d$, $e$, $f$ onto the plane are 120$^\circ$.

If $a \ge 1,~b,~c \ge 0$, then the vertex condition (\ref{vert}) cuts off an equilateral triangle on the plane $d+e+f = 0$.
Its vertices are $(2\mu,-\mu,-\mu)$, $(-\mu,2\mu,-\mu)$, $(-\mu,-\mu,2\mu)$ (denoted $D$, $E$, $F$ respectively). In the function \lstinline{plot} we prepare the $x,y$ coordinates of the points of the triangle as follows:

\begin{lstlisting}
mu = min(a-1,b,c)
r = 6**.5 * mu
xmax = 3**.5/2*r
x = np.linspace(-xmax,xmax,101)
XV = np.concatenate((x,x))
YV = np.concatenate((-r/2*np.ones(x.shape),r-3**.5*np.abs(x)))
\end{lstlisting}

To plot the rest of the listed shapes, we choose the polar coordinate system on the plane centered in the point $d=e=f=0$. We express the Cartesian coordinates $\{d,e,f\}$ by the polar coordinates $\{r,\phi\}$ in the following way:

\begin{equation}
    d = \sqrt{\frac{2}{3}} r \cos \phi,
    \quad
    e =\sqrt{\frac{2}{3}} r \cos \left( \phi + \frac{2\pi}3 \right),
    \quad
    f = \sqrt{\frac{2}{3}} r \cos \left( \phi - \frac{2\pi}3 \right),
\end{equation}
    
Now one can rewrite the edge conditions (\ref{edge1}) - (\ref{edge3}) as:
%this is the same equation, it is enough to write only once for \theta=\phi, phi\pm \frac{2\pi}3
 \begin{align}
        \sqrt{
        \left( a-1- \frac r{\sqrt{6}}  \cos \phi \right)^2
        - 
        \frac{r^2}{2} \sin^2 \phi
        }
        +
        \sqrt{
        \left( \frac{b+c}2 + \sqrt{\frac{2}{3}} r \cos \phi\right)^2
        -
        \left( \frac{b-c}2 \right)^2
        }
        \ge 1, \label{edge1_polar} \\
        \sqrt{
        \left( a-1- \frac r{\sqrt{6}}  \cos \left( \phi + \frac{2\pi}3 \right) \right)^2
        - 
        \frac{r^2}{2} \sin^2 \left( \phi + \frac{2\pi}3 \right)
        }
        +
        \sqrt{
        \left( \frac{b+c}2 + \sqrt{\frac{2}{3}} r \cos \left( \phi + \frac{2\pi}3 \right) \right)^2
        -
        \left( \frac{b-c}2 \right)^2
        }
        \ge 1, \label{edge2_polar} \\
        \sqrt{
        \left( a-1- \frac r{\sqrt{6}} \cos \left( \phi - \frac{2\pi}3 \right) \right)^2
        - 
        \frac{r^2}{2} \sin^2 \left( \phi - \frac{2\pi}3 \right)
        }
        +
        \sqrt{
        \left( \frac{b+c}2 + \sqrt{\frac{2}{3}} r \cos \left( \phi - \frac{2\pi}3 \right) \right)^2
        -
        \left( \frac{b-c}2 \right)^2
        }
        \ge 1. 
        \label{edge3_polar}
    \end{align}

We obtain the same arcs of a bean curve rotated by $120^\circ$, forming a curved-arc triangle.
Let us focus on the arc lying antipodal to the point $D$. It spans between angles $[2\pi/3,4\pi/3]$, but due to its symmetry, it is enough to calculate the first half and then mirror the result to the second half.

To obtain the formula for the arc, we proceed as follows.
For a given $\phi$ we square twice the arcs equation to get rid of square roots and obtain a four-order polynomial. Next, we calculate its roots and compare which of them is positive and satisfies the original equation (before squaring). If there is no such a root or it is outside the blue triangle, we notice it by putting the value \lstinline{np.NaN} for such $\phi$:

\begin{lstlisting}
def satisfy_original(a:float,b:float,c:float,phi:float,r:float,tol:float=1e-6) -> bool:
	res = ((a-1-r*np.cos(phi))**2 - 3*r**2*np.sin(phi)**2)**.5 + ( ((b+c)/2 + 2*r*np.cos(phi))**2 - ((b-c)/2)**2 )**.5 - 1
	return np.abs(res) < tol 

def in_triangle(a: float,b: float,c: float,phi: float,r: float) -> bool:
	return r <= .5* min(a-1,b,c) / max(-np.cos(phi),-np.cos(phi+2*np.pi/3),-np.cos(phi-2*np.pi/3))


def on_roots(a: float,b: float,c: float,phi: float,roots: np.ndarray) -> float:
	roots = [r.real if r.imag == 0 and r>0 and satisfy_original(a,b,c,phi,r) and in_triangle(a,b,c,phi,r) else np.NaN for r in roots] 
	m = np.nanmin(roots)
	return m
	
def arc_polar(a: float,b: float,c: float) -> Callable[[np.ndarray], np.ndarray]:
	def inner(phi: np.ndarray) -> np.ndarray:
		coeff4 = 9/4; coeff4 = coeff4*np.ones(phi.shape)
		coeff3 = 3*(a+b+c-1) * np.cos(phi)
		coeff2 = ((a+b+c-1)**2-4)*np.cos(phi)**2 - 1.5*((a-1)**2-b*c-1)
		coeff1 = -((a+b+c-1)*((a-1)**2-b*c-1)+2*(b+c))*np.cos(phi)
		coeff0 = ((a-1)**2-b*c-1)**2/4-b*c; coeff0 = coeff0*np.ones(phi.shape)
		P = np.array([coeff4,coeff3,coeff2,coeff1,coeff0]).transpose()
		P = np.array([on_roots(a,b,c,phii,np.roots(p))*6**.5 for phii, p in zip(phi,P)])
		return P
	return inner
\end{lstlisting}

In this way we calculate an array of the proper values of $r$ for a range of angles $[\pi/3,\pi]$ - twice wider than the halt of the range of the arc. We do it, because the shape we are looking for will be an intersection of areas bounded by arcs and in general we do not know, if for a given $\phi$, the point of the considered arc, or a point of the prolonged neighbouring arc are closer to the origin. 

Hence, in the function \lstinline{plot} we take the arc prolonged to the twice wider range $[\pi/3,\pi]$, then fold it in the endpoint of the original range, and take the minimum of two values of radius for each $\phi$. In this way we obtain the half of the arc in the considered range. Reflecting it with respect to the value $\pi$, we obtain the arc in the whole range. Next we produce the arrays of $x$-coordinates and $y$-coordinates of points of all three arcs:

\begin{lstlisting}
phi = np.linspace(np.pi/3,np.pi,201);	r = arc_polar(a,b,c)(phi); # twice prolonged half of arc
r = np.nanmin(np.vstack((r[:101][::-1],r[100:])), axis = 0);	   # folding prolongation and taking minimum
r = np.hstack((r,r[:0:-1]))                          # mirror to the second half
phi = np.pi/2 + np.linspace(2*np.pi/3,4*np.pi/3,201) # proper range of phi
XE= np.hstack((r*np.cos(phi), r*np.cos(phi+2*np.pi/3), r*np.cos(phi-2*np.pi/3)))
YE= np.hstack((r*np.sin(phi), r*np.sin(phi+2*np.pi/3), r*np.sin(phi-2*np.pi/3)))
\end{lstlisting}
(the shift of the range of the angle by $\pi/2$ is because the axis $d$ points up.)

Next we generate the arrays of $x$ and $y$ coordinates of points of the green circle bounding the set of points representing operators satisfying the hessian condition:
\begin{lstlisting}
r = (a+b+c-((a-2*b-2*c)**2+3*(b-c)**2)**.5)/6**.5
XP = r*np.cos(np.linspace(0,2*np.pi,601)) 
YP= r*np.sin(np.linspace(0,2*np.pi,601))
\end{lstlisting}

Finally, to calculate the points of the boundary of the set of completely positive matrices, we transform the saturation of inequality (\ref{CP2}) to the form:
\begin{equation}
    def + (a-1) (de+ef+fd) + a^3 - 3a^2 = 0
\end{equation}
in the polar coordinates:
\begin{equation} \label{CP_poly}
    \cos\phi \left( \cos^2\phi -\frac 34 \right) \sqrt{\frac 23} r^3 - \frac {a-1}2 r^2 + a^2(a-3) = 0. 
\end{equation}
Here we proceed in a similar fashion as we described for the case of the red curved-arc triangle. For a  given $\phi$ we calculate the real, positive roots of the polynomial \ref{CP_poly}. If such obtained radial coordinate exceeds the radial coordinate of the point on the triangle, then we take the point on the triangle, because complete positivity implies positivity (the set of completely positive maps is a subset of the set of positive maps). The necessary condition for complete positivity is $a \ge3$, hence otherwise we return an array of \lstinline{np.NaN} (empty plot):
\begin{lstlisting}
def on_CP_roots(roots: np.ndarray, bound: float):
	roots = [(r.real if r < bound else  bound) if r.imag == 0 and r>0 else np.NaN  for r in roots] 
	m = np.nanmin(roots)
	return m

def CP(a:float, b:float, c:float) -> Callable[[np.ndarray], np.ndarray]:
	if a <= 3:
		return lambda phi: np.nan * phi
	def inner(phi: np.ndarray) -> np.ndarray:
		coeff3 = np.cos(phi)*(np.cos(phi)**2-.75) * (2/3)**1.5
		coeff2 = -(a-1)/2 * np.ones(phi.shape)
		coeff1 = np.zeros(phi.shape)
		coeff0 = a**2*(a-3) * np.ones(phi.shape)
		P = np.array([coeff3,coeff2,coeff1,coeff0]).transpose()
		tri = (1.5**.5 * min(a-1,b,c) / max(-np.cos(phii),-np.cos(phii+2*np.pi/3),-np.cos(phii-2*np.pi/3)) for phii in phi)
		P = np.array([on_CP_roots(np.roots(p),bound) for p,bound in zip(P, tri)])
		return P
	return inner
\end{lstlisting}
The above function is then invoked in the function \lstinline{plot} to generate the arrays of $x$ and $y$ coordinates:
\begin{lstlisting}
phi = np.linspace(0,2*np.pi,601)
r = CP(a,b,c)(phi)
phi = np.pi/2 + phi
XC = r*np.cos(phi)
YC = r*np.sin(phi)
\end{lstlisting}

The plots are obtained from the data stored in 1-d arrays \lstinline{XV,YV,XE,YE,XP,YP,XC,YC} returned by the function \lstinline{plot}.

\end{document}